# Reconstruction of Worm Propagation Path Using a Trace-back Approach

Sara Asgari
*Department of Computer Engineering*
*Amirkabir University of Technology*
Tehran, Iran
s.asgari@aut.ac.ir

Babak Sadeghiyan
*Department of Computer Engineering*
*Amirkabir University of Technology*
Tehran, Iran
basadegh@aut.ac.ir

***Abstract*—Worm origin identification and propagation path reconstruction are essential problems in digital forensics. However, a small number of studies have specifically investigated these problems so far. In this paper, we extend a distributed trace-back algorithm, called Origins, which is only able to identify the origins of fast-spreading worms. We make some modifications to this algorithm so that in addition to identifying the worm origins, it can also reconstruct the propagation path. We also evaluate our extended algorithm. The results show that our algorithm can reconstruct the propagation path of worms with high recall and precision, on average around 0.96. Also, the algorithm identifies the origins correctly in all of our experiments.**

## I. INTRODUCTION

Investigation of worms is very important for digital forensics. Knowledge of the origins and the propagation path of worms helps in identifying the nodes which are responsible for entering the worm to the network and specifying the sequence of nodes that got infected in the propagation process. Identifying the origins of worms and reconstructing their propagation path are necessary for evidence collecting and investigating and can help the investigator to know how the attack have occurred in the network and how the network defenses have been violated. They have significance in understanding the cause of risks, identifying network security weaknesses, doing better security measures and mitigating damages. They are also helpful in predicting and mitigating future attacks. Despite the importance of worm origin identification and propagation path reconstruction, so far very few researches have considered these problems. Besides, the existing methods have some limitations. The proposed methods in [1], [2], and[3] assumed that the complete graph of hosts' communications is available. However, in most cases obtaining this graph is very difficult. Shibaguchi et al. method[4] is not fully automated. Tafazzoli et al. method[5] requires a learning process for each worm before the real propagation.

L. Burt et al. (2008)[6] proposed a distributed trace-back approach, called Origins algorithm, to identify the propagation roots of fast-spreading worms. Trace-back is the process of tracing worm to its origins. This algorithm is based on the behavior of the infected nodes, i.e., when a node becomes infected, it immediately tries to identify new victims. Origins utilizes the Intrusion Detection System (IDS) to detect the worm outbreak. Therefore, timely and accurate worm detection by IDS is one of the preconditions of this algorithm. Also, installing the monitor agent on all hosts of the network is another precondition. In Origins, each monitor agent identifies its parent with the eventual goal of the origin identification. Each monitor agent records the incoming connection information within a sliding window and queries every source host in its incoming connection list, when the worm outbreak is detected by IDS, and correlates events to choose its parent. Roots are the nodes that don't have any parent.

In this paper, we add the ability to reconstruct the propagation path of the worms to the Origins algorithm. To do this, we make some modifications to it. Among the nodes which are the candidates for being the parent of a node, the Origins algorithm arbitrarily chooses the first candidate as the parent. However, our extended Origins algorithm determines the parent by considering some conditions.

This paper is organized as follows: Section II introduces the related works of identifying the origins of the worms and reconstructing their propagation path. Section III presents the Origins algorithm and its preconditions. Our proposed algorithm, extended Origins, is described in section IV. Experimental evaluation of our algorithm is presented in section V. Section VI provides a comparison of our algorithm with other worm origin identification and propagation path reconstruction methods. Section VII outlines the conclusions and presents future works.

## II. RELATED WORKS

Although several methods, e.g. [7-18], have been proposed to identify the propagation origins in networks so far, the number of methods that have been specifically proposed to identify the origins and reconstruct the propagation path of worms is very limited.

The first algorithm for identifying the origin and reconstructing the propagation path of worms was the Random Moonwalk algorithm which was proposed by Xie et al. (2005) [1]. In this algorithm, it was assumed that all hosts' communications are available. It reconstructed the propagation path of the worm by walking backward in time along paths of flows.

Xiang et al. (2008)[2] introduced the offline Accumulation algorithm and then proposed the online Accumulation algorithm based on that using the sliding window. Shi et al. (2009)[3] improved the online Accumulation algorithm by using the Bayesian network.

Shibaguchi et al. (2009)[4] proposed a method that combines automatic algorithms and human interactions. First, an automatic algorithm is run on network logs to remove false negatives. Then analysts analyze the result by using a visualized log window to reduce false positives.

To the best of our knowledge, from 2010 to 2018, no research has been reported to address the problem of identifying the origin and reconstructing the propagation path of worms.

Tafazzoli et al. (2019)[5] constructed a probabilistic model that could be used to estimate the infection probability of nodes by receiving some features from the network, and they proposed a back in time algorithm to identify the origin and to reconstruct the propagation path of local preference scanning worms using the learned model.

The Origins algorithm, proposed in [6], can be used to identify the origins of worm propagation. We extend this algorithm to be able to reconstruct the propagation path of the worm, too.

## III. Origins Algorithm

L. Burt et al. [6] proposed an automatic distributed mechanism, called the Origins algorithm, to identify the origins of fast-spreading worms. In this section, we describe this algorithm and its preconditions.

### *A. Preconditions*

- In this algorithm, the IDS performs worm detection. This is not a limitation because nowadays, administrators often utilize IDSs to protect networks against intrusions. However, accurate and timely detection of worm is a precondition of this algorithm[6].
- This algorithm assumes that the worms don't use spoofed source IP addresses. This assumption is close to reality. Worms almost always don't use spoofed source IP addresses to prevent filtering. Also, TCP-based worms can't use spoofed addresses due to the three-way handshake[6].
- The monitor agent is installed on every host. In this paper, we don't consider the situation that the monitor agent is not installed on all hosts. However, this case is discussed in [6].

### *B. Description*

In this algorithm, the worm detection is performed by IDS. In addition, a monitor agent is installed on every host. Each monitor agent maintains the header information of incoming TCP and UDP connections within a sliding window. The size of the sliding window depends on the speed characteristics of the worm. When IDS detects a worm outbreak, it directs the monitor agent to enter the trace-back state. In the query phase, the monitor agent investigates its incoming connection list to extract the source IP addresses and sends a query message to each of them. The host that receives the query message replies with the time of infection if a worm outbreak is detected by IDS. Otherwise, it replies with "No". After receiving replies, the querying monitor agent decides which node is its parent. The node that replies with "Yes" is considered as the parent. If the monitor agent receives "Yes" reply from more than one node, arbitrarily, the first "Yes" reply is chosen, because following any node that answers with "Yes" finally leads back to the origins. The nodes that have no parent are considered as the worm origins[6].

## IV. Our Extended Origins Algorithm

In this section, we describe our proposed algorithm, called extended Origins, that in addition to identifying the origins of the fast-spreading worms, it also reconstructs the propagation paths. This algorithm is based on the Origins algorithm, described in the previous section, and is proposed by making some modifications to it. Note that the preconditions of this algorithm are the same as the preconditions of the Origins algorithm.

In this algorithm, after detecting the worm outbreak by IDS, the node's monitor agent enters the trace-back state. In the query phase, in addition to extracting the source IP addresses of the incoming connection list, the monitor agent extracts the time that each connection is initiated. Then it sends a query message to each of these IP addresses. In the reply phase, each infected node replies with its infection time, and each non-infected node replies with "No". After receiving replies, the querying monitor agent considers the node that replies with "Yes" as its parent. If the querying monitor agent receives "Yes" reply from more than one node, the node that satisfies the following conditions is considered as the parent:

1) the time that the node initiated the connection is greater than or equal to its infection time (it initiated the connection after it became infected).

2) Among the nodes that satisfy the first condition, the node that initiated the connection before others.

Origins are the nodes with no parent.

The propagation path of a worm in a network is represented in Fig. 1. The incoming connections of node 5 within the sliding window is also represented in this figure as an example. In Fig. 1. straight dashed lines illustrate normal connections (connections that don't carry the infectious payload), curved dashed line illustrates unsuccessful attack connection (the connection that carries the worm attack traffic but it doesn't infect the destination host), and solid lines illustrate successful attack connections, i.e., worm propagation path.

Nodes 6, 14, 16, 2, and 3 initiate the connection to node 5 within node 5's sliding window. Within this sliding window, nodes 14 and 16 are non-infected. Therefore, the connections of nodes 14 and 16 are normal connections. Although node 6 is infected within this sliding window, it initiates the connection to node 5 before it becomes infected. So the connection of node 6 is also a normal connection. Within this sliding window, nodes 2 and 3 are infected. In addition, they initiate the connection to node 5 after they become infected, so their connections contain the infectious payload. Although the connection initiated by node 3 is an attack connection, node 5 isn't infected by node 3 because it becomes infected earlier by node 2.

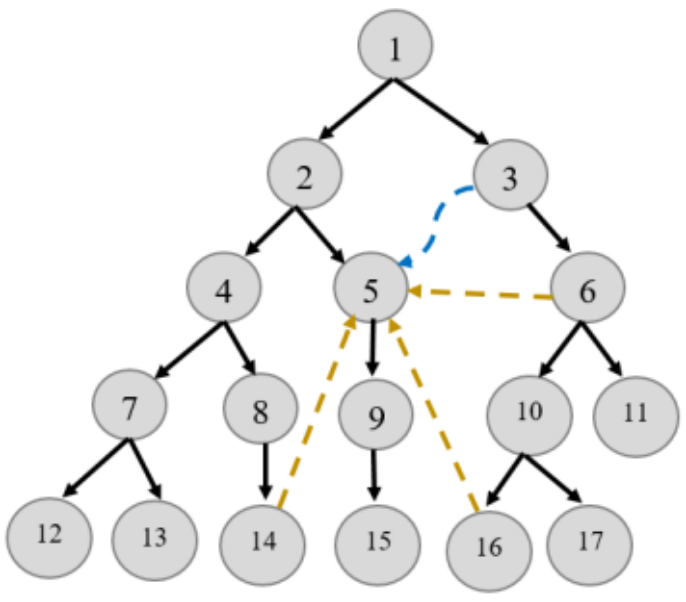

Fig. 1. The worm propagation path in a network.

Fig. 2. represents the trace-back state of node 5. As it is illustrated in Fig. 2., within this node's sliding window, nodes 6, 14, 16, 2, and 3 initiate the connection to this node at times $t_{c6}$, $t_{c14}$, $t_{c16}$, $t_{c2}$, $t_{c3}$, respectively.

In the query phase, node 5's monitor agent extracts the source IP address and the time of initiation of each connection and queries to each of the IP addresses.

In the reply phase, nodes 14 and 16 reply with "No". So neither of these nodes can be the parent of node 5. In addition, nodes 6, 2, and 3 reply with the infection time, $t_{i6}$, $t_{i2}$, and $t_{i3}$, respectively. The time that node 6 initiates the connection to node 5 ($t_{c6}$) is not greater than the infection time of node 6 ($t_{i6}$). So obviously, this connection is a normal connection. Nodes 2 and 3 satisfy the first condition, i.e., the time that they initiate the connection is greater than their infection time, so one of these two nodes infects node 5. Node 2 initiates the connection before node 3, so node 2 is the parent of node 5.

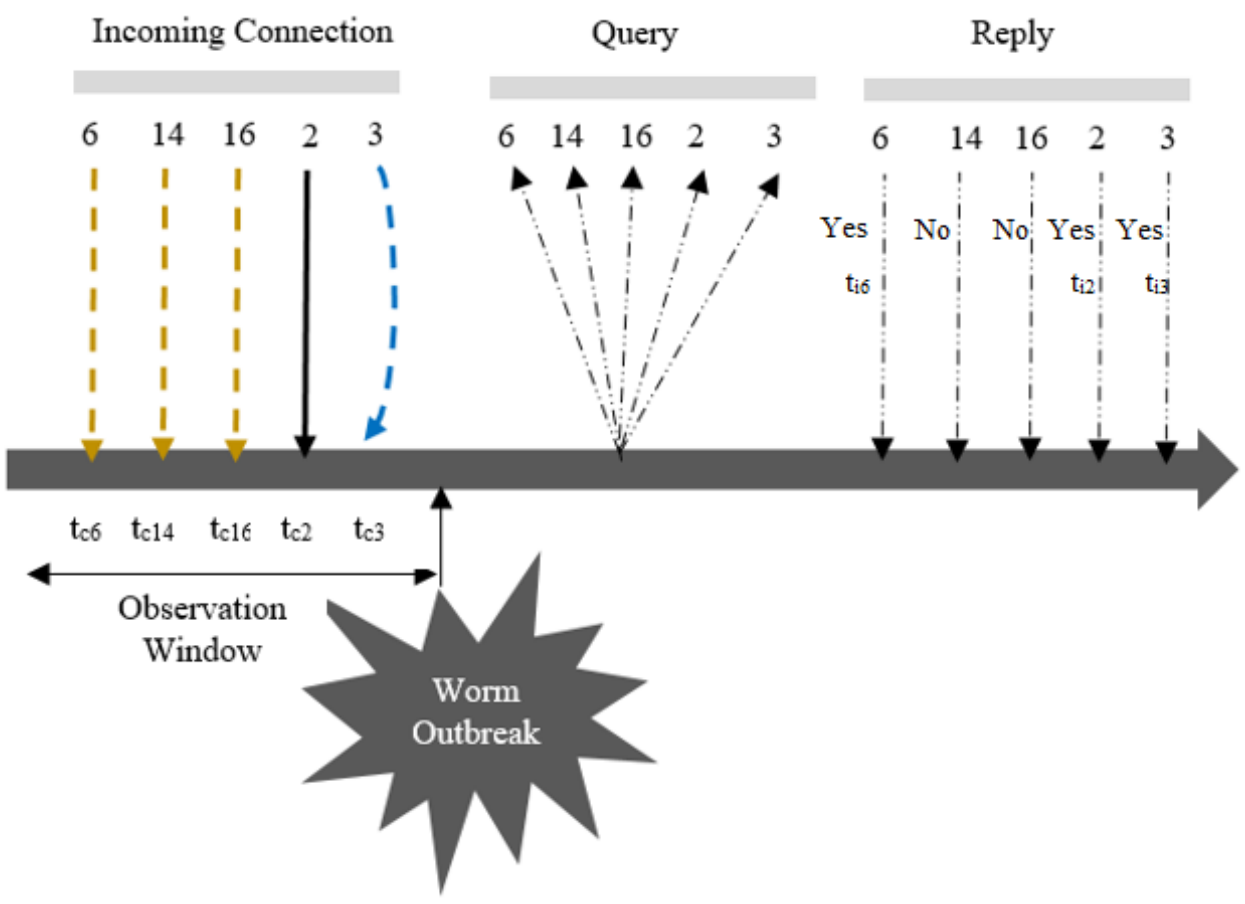

Fig. 2. Trace-back state of node 5.

## V. Experiments

We evaluate our extended Origins algorithm for a number of TCP and UDP scanning worms. In [19], two categories of datasets were generated that each category contains several sets of traffic traces. These datasets contain normal background traffic and worm traffic and are used to evaluate the worm origin identification and propagation path reconstruction methods. The network topology to generate the datasets of category I consists of 200 end-hosts that are located in four subnets. The normal traffic of datasets of category I consists of HTTP, HTTPS, DNS, SSH, FTP, Email, and ping. Category I contains six sets of traffic traces. The simulated worms in sets 1, 4, and 5 are Slammer[20], Code Red I, and Code Red II[21], respectively. The simulated worms in sets 2, 3, and 6 are the modified versions of Slammer and Code Red. We use sets 1, 4, and 5 to evaluate our algorithm. The worm propagation and the infection network parameters in these sets are illustrated in Table I. There are three traffic traces in each set that their worm origins and propagation paths are different.

Slammer was a UDP-based uniform random scanning worm that infects hosts by exploiting a buffer overflow vulnerability in Microsoft's SQL servers. Code Red I and Code Red II were TCP-based worms that exploited a buffer overflow vulnerability in Microsoft's IIS web servers. The scanning strategy of Code Red II was the local preference. However, the Code Red I was a uniform random scanning worm.

To evaluate the recall and precision of this algorithm, we consider three test plans that each of them consists of three experiments. To perform the experiments of the test plans 1, 2, and 3, we use the traffic traces of sets 1, 4, and 5 of category I, respectively. The number of the origins in experiments 1, 2, and 3 of the test plan 2 is 3, 2, and 4, respectively. However, in other experiments, only one node is the origin of the worm propagation. As previously mentioned, the size of the sliding window is chosen according to the speed characteristics of the worm. Correctly determining the size of the sliding window is essential for the high accuracy of our algorithm. Larger window size has no effect on the accuracy of the algorithm. However, determining a smaller value for the size of the sliding window leads to lower accuracy. In all of our experiments, the size of the sliding window is determined 1 second.

The results of the propagation path reconstruction in our experiments, i.e., real propagation path and the reconstructed path, are illustrated in Table III. These results are also summarized in Table II.

For example, in experiment 2 of test plan 3, 26 edges exist in both the real propagation path of the worm and the reconstructed path (true positive). The edge 12_1→35_1 exists in the real propagation path of the worm, but the algorithm fails to detect it in the reconstructed path (false negative). The edge 16_0→35_1 exists in the real propagation path of the worm, but the algorithm reports it in the reconstructed path (false positive), illustrated with the red dotted line in Table III.

In all of our experiments, the origins are correctly identified. Our results are similar to the results of the experiments conducted in [6] on the Origins algorithm. As in [6], the success rate in identifying the origins is 100% when the monitor agent is installed on all hosts.

The precision and the recall of our algorithm in reconstructing the propagation path are 0.96, on average.

TABLE I. THE DATASETS USED IN OUR EXPRIMENTS[19]

| **dataset** | **Worm parameters** | | | | | | | **Infection network parameters** | |
|---|---|---|---|---|---|---|---|---|---|
| | ***name*** | ***Transport layer protocol*** | ***Number of concurrent connection (TCP worms)*** | ***Time between probing packets (UDP worms)*** | ***Scanning strategy*** | ***Preference probability (local preference scanning worms)*** | ***Recovery Probability*** | ***Number of nodes*** | ***Type of nodes*** |
| Category I – Set 1 | Slammer | UDP | - | Uniform(4ms, 8ms) | Uniform random | - | $10^{-4}$ per ms | 30 | HTTP Server, HTTPS Server, and Client |
| Category I – Set 4 | Code Red I | TCP | 23 | - | Uniform random | - | $10^{-4}$ per ms | 28 | HTTP Server |
| Category I – Set 5 | Code Red II | TCP | 25 | - | Local Preference | $\frac{1}{8}$: random<br>$\frac{4}{8}$: same class A<br>$\frac{3}{8}$: same class B | $10^{-4}$ per ms | 28 | HTTP Server |

TABLE II. THE RESULTS OF OUR EVALUATIONS

| **Test Plan #** | **Experiment #** | **Path Reconstruction** | | | | | **Origin Identification** |
|---|---|---|---|---|---|---|---|
| | | ***True Positive*** | ***False Negative*** | ***False Positive*** | ***Precision*** | ***Recall*** | |
| Test Plan 1 | Experiment 1 | 29 | 0 | 0 | $\frac{29}{29+0}=1$ | $\frac{29}{29+0}=1$ | ✓ |
| | Experiment 2 | 29 | 0 | 0 | $\frac{29}{29+0}=1$ | $\frac{29}{29+0}=1$ | ✓ |
| | Experiment 3 | 29 | 0 | 0 | $\frac{29}{29+0}=1$ | $\frac{29}{29+0}=1$ | ✓ |
| Test Plan 2 | Experiment 1 | 26 | 1 | 1 | $\frac{26}{26+1}=0.96$ | $\frac{26}{26+1}=0.96$ | ✓ |
| | Experiment 2 | 25 | 2 | 2 | $\frac{25}{25+2}=0.92$ | $\frac{25}{25+2}=0.92$ | ✓ |
| | Experiment 3 | 27 | 0 | 0 | $\frac{27}{27+0}=1$ | $\frac{27}{27+0}=1$ | ✓ |
| Test Plan 3 | Experiment 1 | 27 | 0 | 0 | $\frac{27}{27+0}=1$ | $\frac{27}{27+0}=1$ | ✓ |
| | Experiment 2 | 26 | 1 | 1 | $\frac{26}{26+1}=0.96$ | $\frac{26}{26+1}=0.96$ | ✓ |
| | Experiment 3 | 24 | 3 | 3 | $\frac{24}{24+3}=0.88$ | $\frac{24}{24+3}=0.88$ | ✓ |

TABLE III. THE PROPAGATION PATH RECONSTRUCTION RESULTS IN OUR EXPERIMENTS

| **Test Plan #** | **Path** | **Experiment #** | | |
|---|---|---|---|---|
| | | ***Experiment 1*** | ***Experiment 2*** | ***Experiment 3*** |
| Test Plan 1 | Real path | 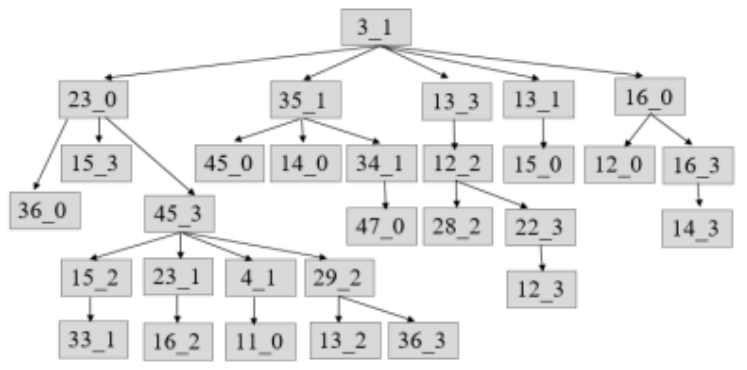 | 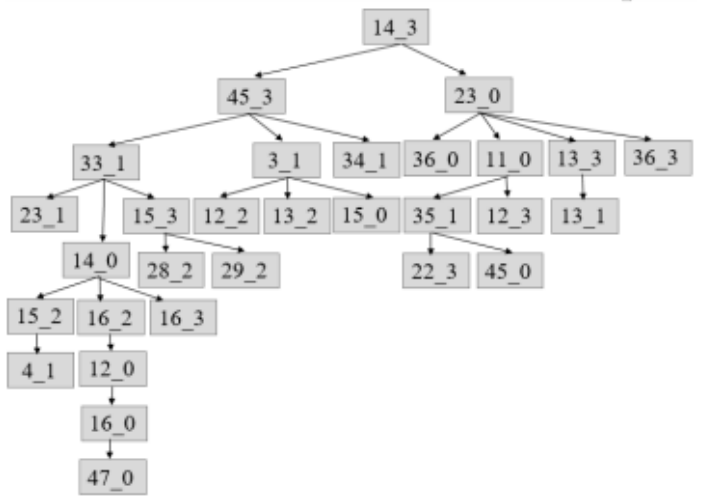 | 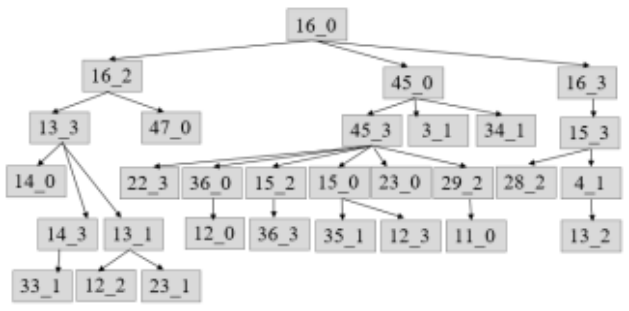 |
| | Reconstructed path | 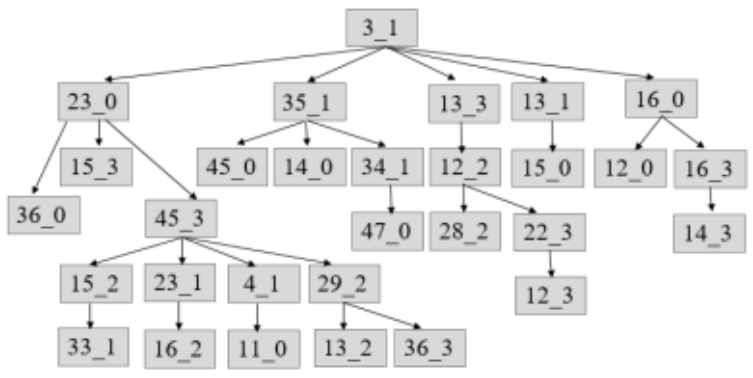 | 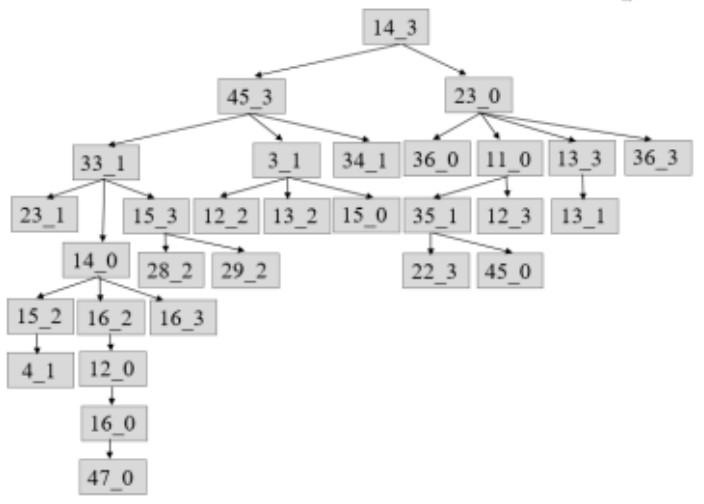 | 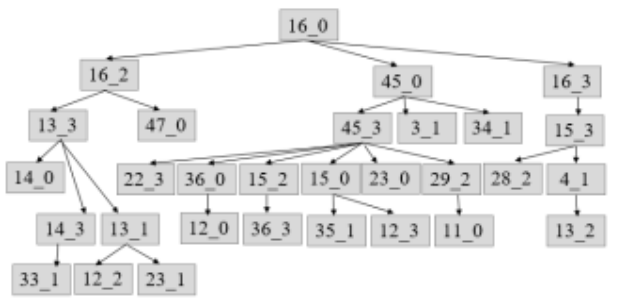 |
| Test Plan 2 | Real path | 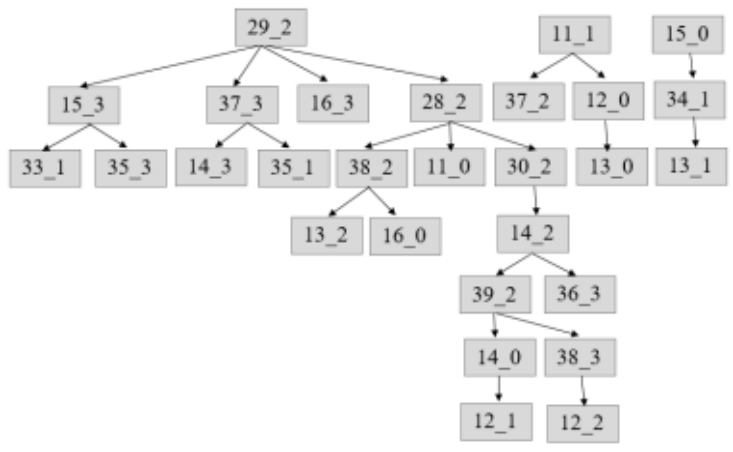 | 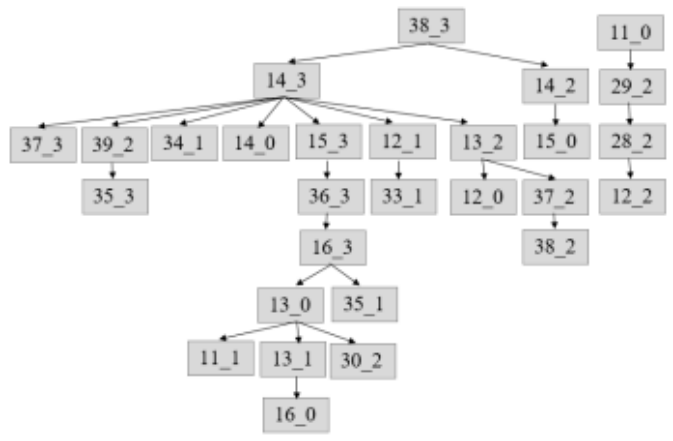 | 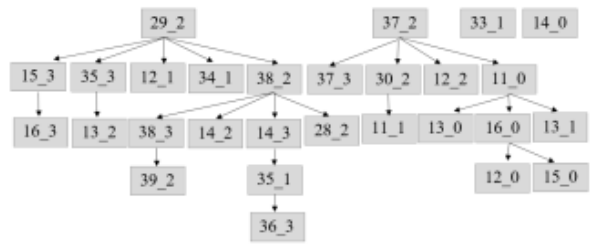 |
| | Reconstructed path | 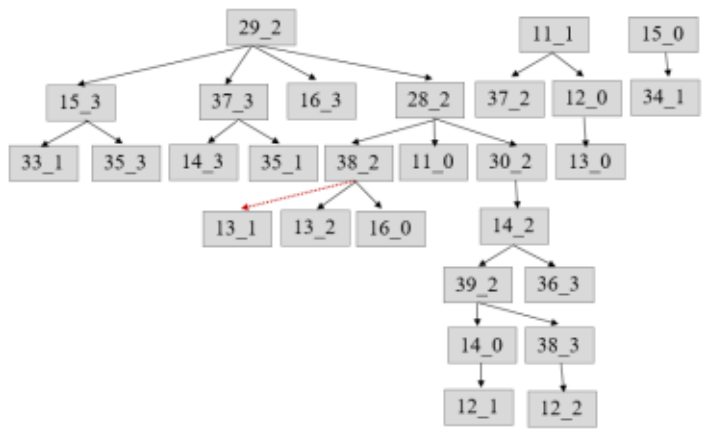 | 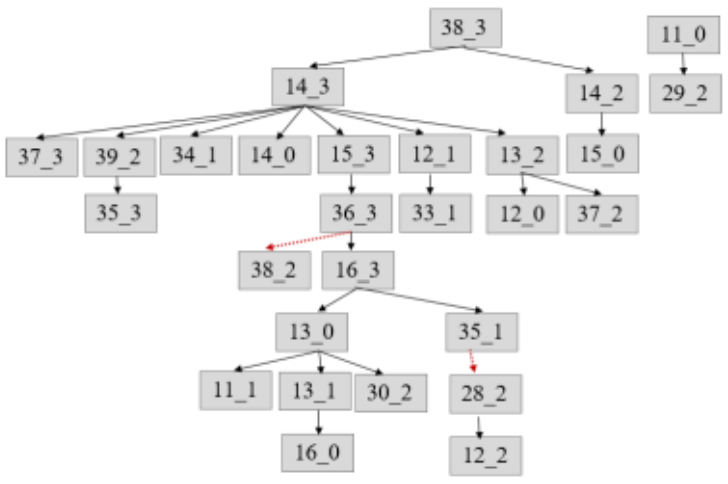 | 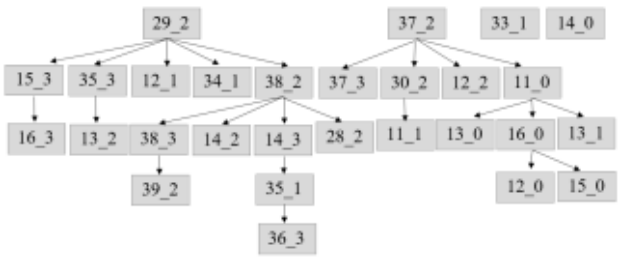 |
| Test Plan 3 | Real path | 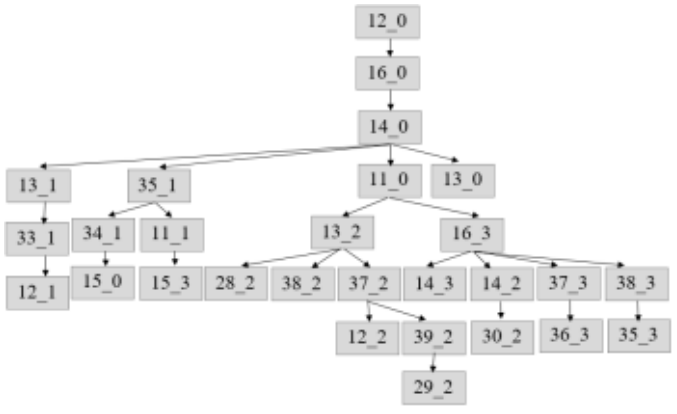 | 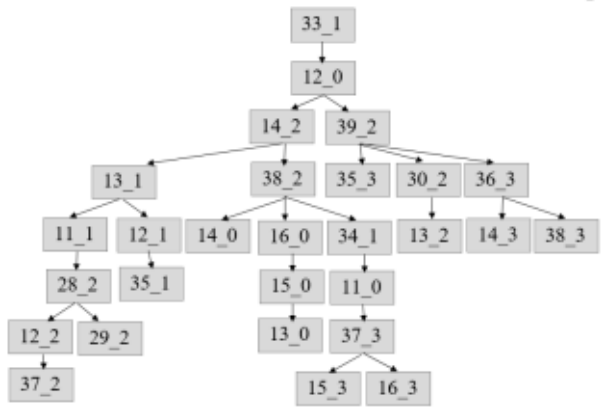 | 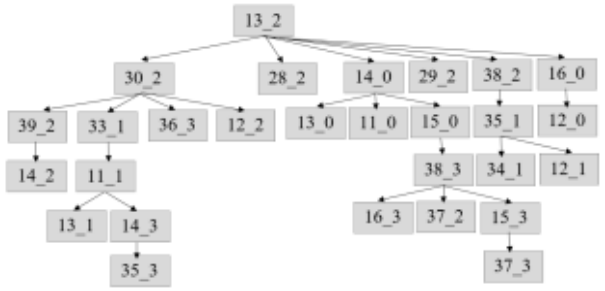 |
| | Reconstructed path | 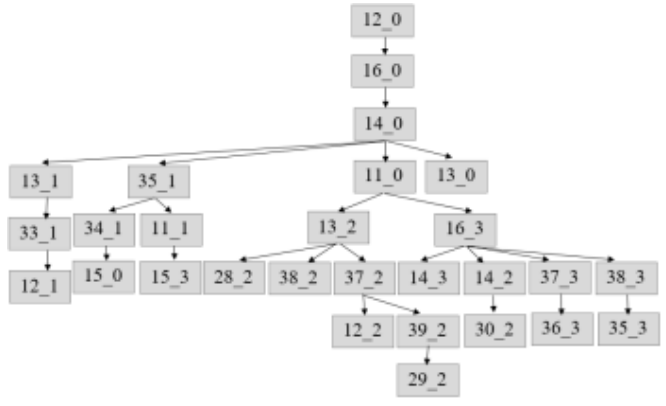 | 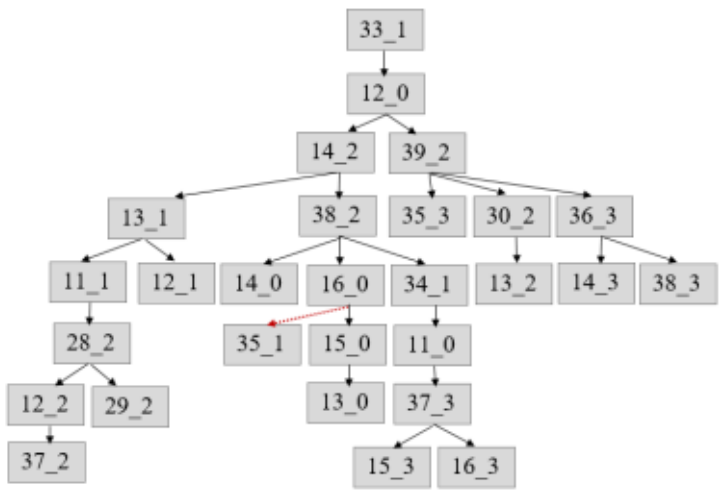 | 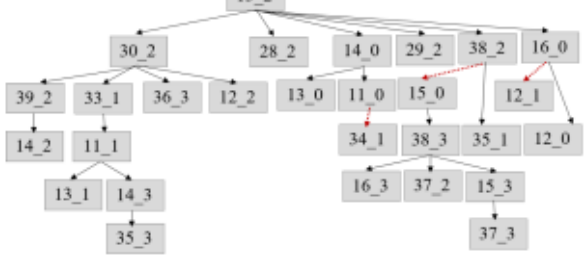 |

## VI. Comparison with Other Methods

In Section II, we introduced the previous works on identifying the origins and reconstructing the propagation path of worms. In this section, we compare our extended Origins algorithm with these methods. These methods have several limitations:

- Xie et al.[1], Xiang et al.[2] ,and Shi et al.[3] methods: They are based on some limiting assumptions. First, only one node is the origin of the worm propagation. Second, the complete graph of hosts' communications is available. Third, worm propagation occurs in a tree-like network.
- Shibaguchi et al.[4] method: it needs the investigations and analyses of analysts and is not fully automated.
- Tafazzoli et al.[5] method: First, it requires a central node to reconstruct the worm propagation path. The central node is a single point of failure for this method. If it fails, this method will not work properly. Second, it requires learning the probabilistic model using the training dataset before the worm propagates in the network. Generating this training dataset, containing both normal background traffic of that network and the propagation traffic of a specific worm, is a challenging task. Third, it is only proposed for a small subset of worms, i.e., local preference scanning worms.

Our extended Origins algorithm does not have any of the limitations of the previous methods mentioned above.

## VII. Conclusion

Origins is a distributed trace-back algorithm proposed to identify the sources of fast-spreading worms. In this paper, we extended this algorithm, called extended Origins, to reconstruct the propagation path too. We also evaluated the extended version for some TCP-based and UDP-based scanning worms, i.e., Slammer, Code Red I, and Code Red II. Our experiments demonstrated that the algorithm is effective and can identify the origins of the worm and its propagation path with good accuracy.

As previously mentioned, one of the preconditions of our extended Origins algorithm is the installation of the monitor agent on all hosts of the network. One of the future works is to improve this algorithm so that the worm propagation path can be reconstructed with acceptable accuracy by installing the monitor agent only on some network hosts.